\documentclass[twocolumn,showpacs,superscriptaddress,preprintnumbers,amsmath,amssymb]{revtex4}
\usepackage{graphicx}% Include figure files
\usepackage{dcolumn}% Align table columns on decimal point
\usepackage{bm}% bold math
\usepackage{lmodern}
\newcommand{\etal}{\textit{et al.}}

\begin{document}

%\preprint{PRL-Preprint}

%\title{n-Propanol on Graphite: From Van-der-Waals Interaction to Hydrogen-Bridge Bonding dominated Multilayer Physisorption}
\title{Transition from van-der-Waals to H-Bonds dominated Interaction\\ in n-Propanol physisorbed on Graphite}

\author{Matthias Wolff}
\affiliation{Experimental Physics, Saarland University, D-66041 Saarbr\"ucken (Germany)}

\author{Frank Kruchten}
\affiliation{Experimental Physics, Saarland University, D-66041 Saarbr\"ucken (Germany)}

\author{Patrick Huber}
\affiliation{Experimental Physics, Saarland University, D-66041 Saarbr\"ucken (Germany)}

\author{Klaus Knorr}
\affiliation{Experimental Physics, Saarland University, D-66041 Saarbr\"ucken (Germany)}

\author{Ulrich G. Volkmann}
\affiliation{Pontificia Universidad  Católica de Chile, Facultad de Física, Santiago (Chile)}
% Pontificia Universidad  Católica de Chile, Facultad de Física, Santiago, Chile


%Ellipsometers, 07.60.Fs
%Films:
%graphene, 68.65.Pq
%growth, deposition, 81.15.-z
%liquid films, 68.15.+e
%Graphite, 81.05.uf
%Physisorption, 68.43.-h

%\author{U. Volkmann}
%\email{}
%\affiliation{Chile}

%\author{K.Knorr}
%\email{ph13kk@rz.uni-sb.de}
%\homepage{http://www.uni-saarland.de/knorr}
%\affiliation{Technische Physik, Universit\"at des Saarlandes\\
%66041 Saarbr\"ucken, Germany}

\date{\today}

\begin{abstract}
Multilayer sorption isotherms of 1-propanol on graphite have been measured by means of high-resolution ellipsometry within the liquid regime of the adsorbed film for temperatures ranging from $180$ to $260$~K. In the first three monolayers the molecules are oriented parallel to the substrate and the growth is roughly consistent with the Frenkel-Halsey-Hill-model (FHH) that is obeyed in van-der-Waals systems on strong substrates. The condensation of the fourth and higher layers is delayed with respect to the FHH-model. The fourth layer is actually a bilayer. Furthermore there is indication of a wetting transition. The results are interpreted in terms of hydrogen-bridge bonding within and between the layers.
\end{abstract}

%\pacs{68.43.-h, 68.08.Bc}

\maketitle

Adsorption and wetting is controlled by the strength and range of the attractive interaction between the admolecules and the substrate in relation to the intermolecular interaction \cite{a,n,b,Bruch2007}. A large amount of pertinent information exists for small molecules on graphite where both interactions are basically of the van-der-Waals (vdW) type, with the substrate interaction being dominant (because of the high polarizibility of graphite that stems from the $\pi$ electrons). On such "strong" substrates the adsorption isotherms show well defined steps indicating layer-by-layer growth that eventually leads to a film of macroscopic thickness at bulk coexistence ("complete wetting") \cite{c}. Consecutive monolayers condense at layering pressures $p_{\rm i}$ ($i$ is the layer index) that reflect the spatial dependence of the substrate vdW potential that decreases with the inverse cube of the distance $d$ from the substrate. Every individual monolayer can be understood as thermodynamic ensemble in two dimensions \cite{o,p} with a phase diagram that includes a gas, a liquid, and one or more solid phases. This view applies in particular to pentane \cite{d}, a molecule similar in size and shape to the n-propanol molecule of the present study.

\begin{figure}
\includegraphics[width=200pt, angle=0]{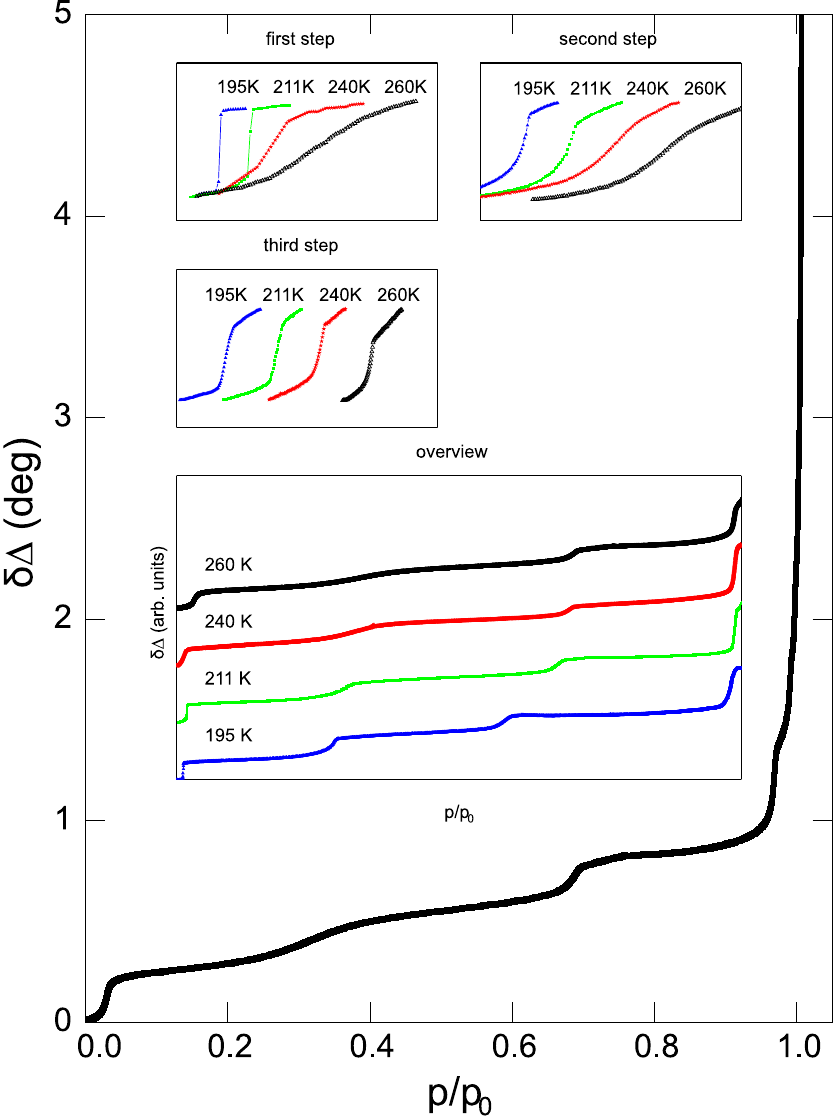}
\caption{(color online). The 260K-sorption isotherm of n-propanol on graphite (ellipsometric coverage $\delta\Delta$ vs. reduced vapor pressure $p/p_{\rm 0}$). The insets show details at $195$K, $211$K, $240$K, and $260$K.}
\label{figA}
\end{figure}

This need not be so for molecules such as water and the alcohols that mutually interact via H bonds but still rely on the vdW interaction with the graphite substrate. Water in fact does not even form a single monolayer on graphite \cite{e}. The existence of H-bonds in solid monolayers of some shorter alcohols \cite{f,g}, including n-propanol \cite{h,i,q}, on graphite has been verified in diffraction studies. The molecules are arranged in such a way that the H-bonds form O-H$\cdot\cdot$O chains that connect neighbouring molecules, quite similar to the situation in the bulk crystals. In the liquid state short segments of such chains persist in spite of considerable conformational disorder, as has been shown e.g. by diffraction on bulk n-propanol \cite{j} and MD simulations on an ethanol monolayer \cite{g}. The H-bonds are also believed to be responsible for some anomalous properties of alcohol monolayers on graphite. The melting temperature of the ethanol monolayer is higher than that of the bulk substance \cite{g}. The n-propanol monolayer melts via a smectic-like mesophase \cite{h}. For tert-butanol the first adsorbed layer is actually a bilayer \cite{k}.

Practically nothing is known, however, about the multilayer regime of H-bonded adsorbed films. Isotherms taken on substrates with large internal surface areas such as exfoliated graphite are faced with the problem of capillary condensation occurring in the cavities and cracks within these substrates. In such experiments the monolayer steps beyond the first one are usually smeared out \cite{u} or are even screened completely. This problem is overcome by measuring the adsorption on the planar surface of compact substrates such as highly oriented pyrolytic graphite (HOPG) using ellipsometry as a surface sensitive probe. In fact this combination has proven to be an excellent tool for the study of multilayer adsorption isotherms \cite{c,l}. Ellipsometric multilayer isotherms on HOPG are of unsurpassed quality. For temperatures $T$s well below the layer critical temperatures such isotherms show almost vertical monolayer steps and almost horizontal plateaus in between. We have applied this technique to n-propanol on HOPG. The multilayer adsorption of pentane \cite{d} will serve as vdW reference.

The ellipsometric observable of interest is $\delta\Delta$. $\Delta$ is the phase shift between the electric field components parallel and perpendicular to the reflection plane introduced by the reflection from the surface. $\delta\Delta$ is the change of $\Delta$ upon film adsorption. $\Delta$ is measured with a resolution of $10^{-3}$~deg. The signal expected for one monolayer can be estimated from a microscopic model \cite{m}, the relevant input parameters of the adsorbate being the molecular polarizibility and the coverage $\Theta$ (= number of molecules per area). For propanol and our ellipsometric setup one arrives at $\delta\Delta=0.29$~deg for a monolayer of molecules lying flat on the substrate. This value is based on X-ray diffraction data on the coverage of the solid monolayer \cite{h}. For a hypothetical tight packing of perpendicular molecules a value of $0.42$~deg is estimated. Since the polarizibility of propanol is almost isotropic, the difference for the two $\delta\Delta$ values is almost entirely due to the different coverages $\Theta$ of the two orientations.
    
The nominal impurity of n-propanol used is 99.5\%. The graphite sample is embedded in a Cu block which has bore holes for the passage of the incoming and the reflected light beam of the ellipsometer (see inset in Fig.~2). The Cu block consists of two parts, bolted together with the graphite sample in between. The design of the UHV/cryogenic setup is such that the Cu block is the coldest part that is in contact with the propanol vapor. A monochromator grade HOPG sample with a mosaic width about the surface normal of $0.4\pm0.1$~deg has been used. Sorption cycles have been performed by introducing and withdrawing the vapor through a dosing valve. A cycle takes typically three hours, a compromise between fast operation upon which global adsorption/desorption hysteresis appears and slow operation upon which impurities accumulate in the UHV chamber. 20 isotherms have been measured, covering $T$s from $180K$ to $260K$ and vapor pressures from $10^{-5}$ to $10$~mbar. The triple point of bulk propanol is at $149$~K. %Hence the isotherms refer to the liquid regime.     

\begin{figure}
\includegraphics[width=220pt]{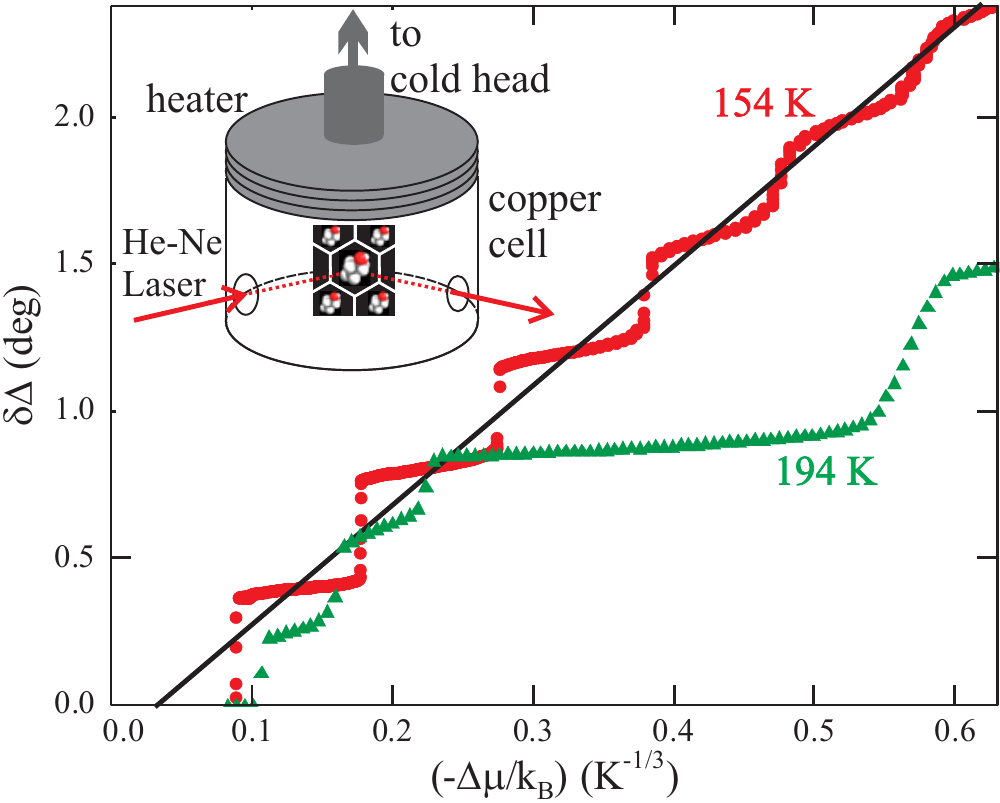}
\caption{(color online). Comparison of the 194K-isotherm of propanol (triangles) and of an 154K-isotherm of pentane (circles), both adsorbed on HOPG and plotted as function of $\Delta\mu^{-1/3}$.}
\label{figB}
\end{figure}

Fig.\ref{figA} displays a selection of isotherms ($\delta\Delta$ as function of the reduced vapor pressure $p/p_{\rm 0}$). The isotherms show four resolved steps and at least at higher $T$ the indication of a fifth one. In the following we discuss the position, height, and shape of the steps. In Fig.~\ref{figB} the $194K$-isotherm is replotted with the reduced vapor pressure being converted into $\Delta\mu^{-1/3}$ where $\Delta\mu=k_{\rm B} T \ln{p/p_{\rm0}}$ is the difference of the chemical potentials of the adsorbed film and of the bulk liquid. For the vdW system pentane/graphite (included for comparison) such a plot shows a regular series of steps of almost identical height and width, demonstrating that i) the partial coverages $\Theta_{\rm i}$ of the individual monolayers are identical and that ii) the chemical potentials of the layers differ just by the amount expected from the $d^{-3}$ decay substrate potential (which means that pentane on graphite conforms to the FHH-model for adsorption \cite{r}). 

For propanol this type of behavior is approximately observed for the first three monolayers, only, suggesting that in first approximation these monolayers are thermodynamically equivalent and that the interaction with the substrate varies again with $d^{\rm -3}$. Note that this law holds not only for the vdW interaction but also for the interaction of the propanol dipole moment with its image in the electrically conducting graphite substrate.

The step heights, taken between the midpoints of the plateaus, of the first three monolayers are about identical, $\delta\Delta_{\rm i}=0.3$~deg. A comparing to the $\delta\Delta$-estimates from above suggests that on average the molecules residing in any of the first three layers lie more or less flat on the substrate, perhaps with some reduction of the "foot print" due to thermal agitation and conformation.
  
The layering pressures can be converted into the $\Delta\mu$-values. One obtains (-45$\pm$5) for the third, (-130$\pm$10) for the second, and $\sim$ -800 for the first layer (in units of $k_{\rm B}$ K). Within experimental error these values are independent of $T$. This means that the partial entropy of these layers is close if not identical to that of the bulk liquid.     

The fourth layer does not form at the vapor pressure estimated from the extrapolation of the FHH-growth but at a considerably higher value that is already close to the saturated vapor pressure $p_{\rm 0}$ (see Fig.~\ref{figB}). The ellipsometric step height of this layer (and also of the fifth one, whenever visible) is about $0.6$~deg. This is more than the estimate for a monolayer of perpendicular molecules, therefore we think of a bilayer. Thus there is a break in the multilayer growth after the adsorption of the first three monolayers. The coupling of the fourth layer to the substrate is weaker than expected from the extrapolation of the $d^{-3}$ dependence and/or the binding within the fourth layer is stronger than for the first three monolayers. The adsorption of the first three layers is dominated by the interaction with the substrate, the condensation of the fourth layer is delayed by enhanced interactions within this layer. 
      
For a vdW multilayer adsorbate on a "strong" substrate, such as pentane on graphite, the individual monolayers are independent from one another. Typically they have quite similar coverage-$T$ phase diagrams with a 2D gas - 2D liquid coexistence region that terminates in a critical point. Isothermal crossing of such a coexistence region leads to quasi-vertical steps of the adsorption isotherms. Thus the experimental quantity to pay attention to is the slope of the isotherms ($\partial\delta\Delta)/\partial\mu$ which is a measure of the isothermal 2D compressibility $\kappa$. 

Does this scenario also apply to the first three layers of propanol on graphite? Fig.~\ref{figA} shows the adsorption steps of the layers 1, 2, and 3 for several $T$s. The first monolayer shows a clear change from quasi-vertical steps at low $T$ to S-shaped steps at higher $T$. A plot of $\kappa^{\rm -1}$ vs. $T$ suggests a critical temperature $T_{\rm c1}$ of this layer of $(209\pm4)$K. (The quality of the data does not allow a reliable derivation of the critical exponent $\gamma$ of $\kappa$ in the hypercritical regime, but within the scatter of the data points the results are compatible with the value of the 2D Ising model, $\gamma=7/4$). Thus there is evidence for a coexistence of a 2D gas-like state and a denser (liquid or solid) 2D state for $T<T_{\rm c1}$. This is in disagreement with the phase diagram proposed in \cite{h,i,q} - see footnote \footnote{Monolayer/submonolayer $\Theta$,$T$ phase diagrams usually show 2D gas/2D solid and 2D gas/2D liquid coexistence regions separated by triple line melting. The diagram proposed in \cite{h,i,q} is not of this type, it rather shows a broad single-phase-region termed "2D solid" with the $T$-melting varying with $\Theta$. The diffraction patterns on two coverages within this phase (that differ by almost a factor of 2) give practically identical Bragg angles \cite{i}. This is in conflict with the idea of a single-phase-region in which the cell area should be proportional to $\Theta^{\rm -1}$.}.

\begin{figure}
\includegraphics[width=200pt, angle=0]{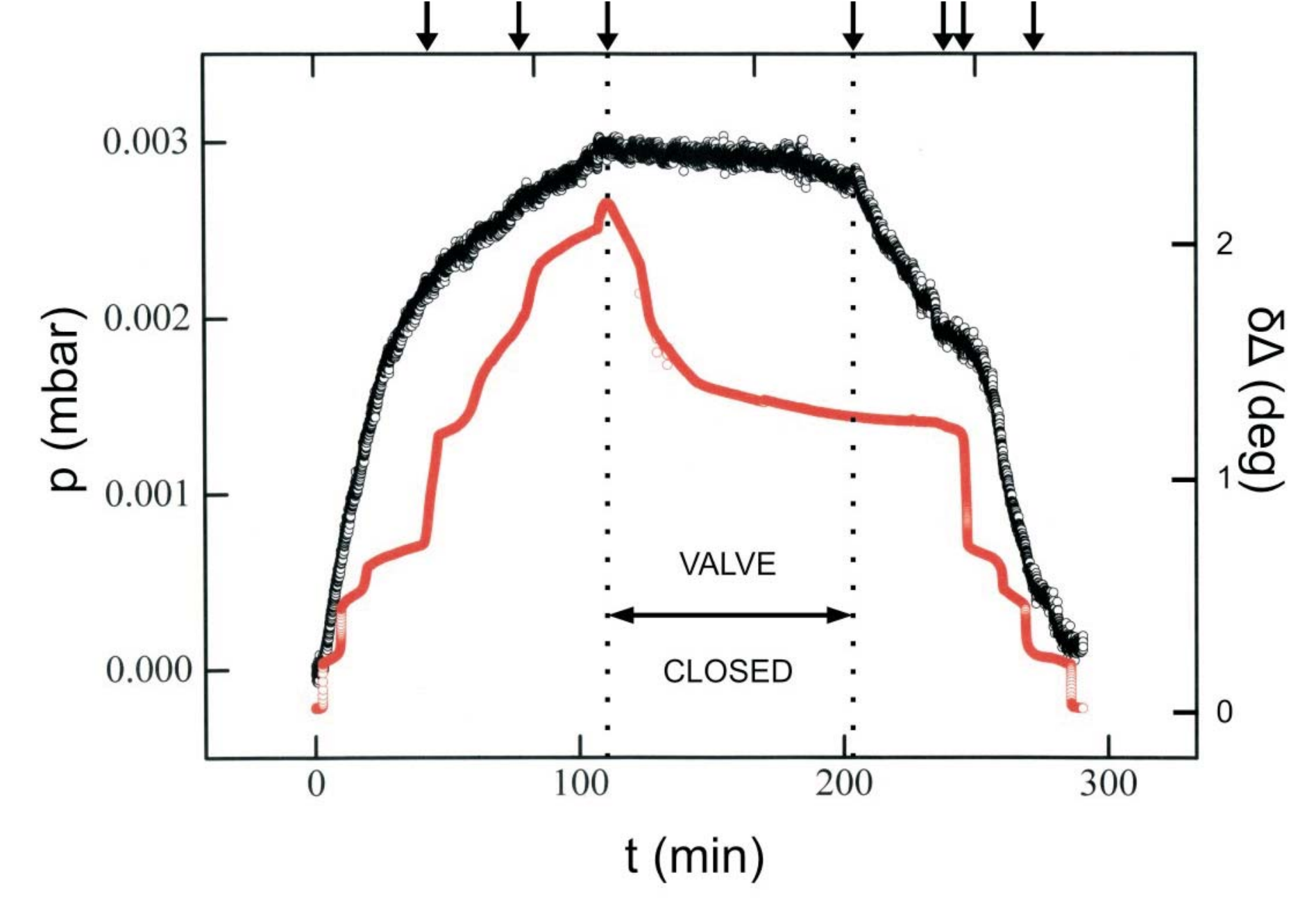}
\caption{(color online). Protocol of an adsorption-desorption cycle at $194$~K showing the time $t$ dependence of the vapor pressure $p$ (black) and of the ellipsometric coverage $\delta\Delta$ (red). Arrows indicate changes of the setting of the dosing valve. In the intermediate period, indicated by vertical dashed lines, the dosing valve is closed. %The two symbols shown for $p$ refer to data obtained with a capacitive membrane gauge and with a spinning ball viscosimeter.
}
\label{figC}
\end{figure}

Analogous coexistence regions are absent in the second and third monolayer. There are no quasi-vertical sections extending over larger parts of these steps (Fig.~\ref{figC}). Furthermore the plateaus between the steps have slopes considerably steeper compared to what has been observed for pentane. Thus one wonders whether one can still refer to discrete layers. Ellipsometric sorption isotherms do not give access to the partial coverages $\Theta_{\rm i}$ of the individual monolayers but just to the total coverage accumulated. Thus the second and third step of the present adsorption isotherms may not exclusively be due to the population of the last layer on top of completed layers underneath (as is the case for vdW systems at $T$s well below the layer critical points), but also to a reorganisation of the total amount of material adsorbed so far, whereupon the distinction of discrete layers is in a multilayer stack is washed out to some extent. On the other hand the existence of residual steps demonstrates that plateau states are less compressible, more reluctant to adsorb additional molecules than states within the steps. 
             
The kinks at the upper end of the second and third step are a feature that deserves special interest. The kinks show up most clearly at low $T$ in the second step and at high $T$ in the third step. The crossover occurs at about $210$K which is just $T_{\rm c1}$. This coincidence suggests some coupling between the first and the higher layers. Just below the kink the isotherms are close to vertical. For the second step at $194$~K, $\kappa$ is reduced by a factor of about 40 from below the edge to above the edge. 

In principle a kink singularity is the signature of a second order phase transition that takes the film from a low-coverage compressible state into a high-coverage state with reduced 2D compressibility. Assuming that the transitions occurring in the second and third step are equivalent, the jump of the transition to higher $\Theta$ and $p$ at higher $T$ appears plausible. At higher $T$ there is more thermal agitation, and hence it needs a larger spreading pressure $\Pi$ to stabilize a high-density phase. $\Pi(p)$ is proportional to $\int^{\rm p}_{\rm 0}\Theta(p) d \ln{(p)}$. For a crystalline film such transitions do not present a conceptual problem. Transitions in the liquid regime are however surprising. In 3D, a transition from a low- to a high-density phase of water, both in the liquid and glassy state, is under active discussion, the two phases being distinguished by different local arrangements of the H-bonds \cite{s}. This view may also apply to the present system, the density and the hydrostatic pressure of the bulk system being replaced by their 2D counterparts, $\Theta$ and $\Pi$. 

Beyond the third step propanol/graphite has turned from a strong into a weak adsorbate/substrate system. Thus one wonders whether the wetting at bulk coexistence is complete or incomplete, and whether there is even a wetting transition. In terms of equilibrium states, the criterion for the character of wetting is straightforward. If the $p\rightarrow p_{\rm 0}$ limit of $\Theta(p)$ corresponds to macroscopic thickness of the adsorbed film, wetting is complete, otherwise incomplete. Unfortunately the experiment gets to its limits in this $p$-range. There is a lot of propanol in the UHV chamber not only in the form of vapor, but also as (capillary) condensate on the rough surface, in the slits and screw holes of the sample holder. The system responds to changes of $p$ induced by readjustments of the dosing valve by slow distillation processes. Furthermore the value of $p_{\rm 0}$ is a priori unknown. Nevertheless there are clear changes between high and low $T$s. For $T>220$~K the sorption isotherms eventually shoot up almost vertically (see the $260$~K-isotherm of Fig.~\ref{figA}), as observed previously for vdW adsorbates on graphite, and we identify the pressure at which this happens with $p_{\rm 0}$. After closing the gas inlet in such a situation, $\delta\Delta$ stays high. For $T<200$~K the behavior is different. This is illustrated in Fig.~\ref{figC}. Even in this $T$-range, $\delta\Delta$ can be driven to values far beyond the fourth step, but after stopping the gas inlet the supersaturated excess coverage evaporates and $\delta\Delta$ decays to a value slightly above the fourth step. We regard this value as the asymptotic thickness of the adsorbed film. This suggests that wetting is incomplete for $T<200$~K and that there is a wetting transition, but at least a thin-thick transition between $200$~K and $220$~K. Between $200$~K and $220$~K, $\delta\Delta$ drifts slowly after closing the gas inlet without settling at a constant value within the duration of the experiment. Thus we cannot give a more precise value of the presumed wetting temperature $T_{\rm w}$. Such slow kinetics in the vicinity of a wetting transition has also been observed on liquid substrates \cite{t}. 
%(A study of the contact angle across $T_{\rm w}$ would be interesting, but the mosaicity of the substrate may interfere with the measurement of very small contact angles.)     
  
The unusual adsorption behavior of propanol on graphite in the multilayer regime is related to the competition of different interactions. They differ not only in strength and range, but also to the extent to which they are directional. We propose the following scenario. At low coverages, the substrate potential dominates. It forces the propanol molecules of the first three monolayers into a flat orientation, which maximizes both the vdW and the dipolar part of the admolecule-substrate binding energy. The sequence of the layering pressures $p_{\rm i}$, $i=1,2,3$, reflects the variation of this energy with the distance from the surface. For submonolayer coverages and lower $T$, one observes the usual coexistence of monolayer puddles and bare regions. The H-bonds within the liquid first monolayer are likely to be a short-range variety of the zigzag O-H$\cdot\cdot$O chains that have been observed in the solid state of the monolayer \cite{h,i,q}. Adding the second and third monolayer on top leads to some re-arrangement of the whole stack because e.g. of a small fraction of interlayer H-bonds that blurs the distinction of individual monolayers, but on the whole the first three layers are very similar, with the molecules oriented parallel to the substrate. After completion of the three-layer-stack the situation changes substantially, the substrate potential has already decayed to a small value and the H-bonds between the molecules are now the dominant interaction. Accordingly the layering pressure of the fourth layer is close to the saturated vapor pressure of the bulk condensate. The molecules of the fourth layer presumably form short range ordered double layers connected by H-bonds.
 
These findings, along with our observation that propanol/graphite may be one of the rare examples of a wetting transition that is not bound to a triple point of the bulk condensate, shed new light on the dominating interactions of an H-bond liquid with a graphitic interface and the resulting near-interfacial structure. Because of the tremendous interest in the physical properties of single-layer graphite, that is graphene \cite{Geim2007}, and adsorption properties on carbon nanotube structures \cite{Rawat2009} we hope that our study will also stimulate further investigations aimed at a more detailed understanding of the rather complex multilayer sorption behavior presented here.
%\acknowledgments
%Such double layers are the leading structural motif of the 3D crystal structures of n-alcohols. 

The authors acknowledge support by the FONDECYT grant no. 1100882.

\end{document}